\documentstyle[aps,prl,psfig,floats]{revtex}
\begin{document}
\draft
\twocolumn[\hsize\textwidth\columnwidth\hsize\csname@twocolumnfalse\endcsname

\title{
Dynamical-charge neutrality at a crystal surface
}
\author{Alice Ruini,$^{1,2}$ Raffaele Resta,$^{1,3}$ and 
Stefano Baroni$^{1,2,4}$}

\address{$^1$INFM -- Istituto Nazionale di Fisica della Materia \\
$^2$SISSA -- Scuola Internazionale Superiore di Stud\^\i\ Avanzati, Via
Beirut 4, 34014 Trieste, Italy \\ $^3$Dipartimento di Fisica Teorica,
Universit\`a di Trieste, Strada Costiera 11, 34014 Trieste, Italy \\
$^4$CECAM -- Centre Europ\'een de Calcul Atomique et Mol\'eculaire, 46 All\'ee
d'Italie, 69007 Lyon, France}

\date{January 1997}
\maketitle

\begin{abstract} For both molecules and periodic solids, the ionic dynamical
charge tensors which govern the infrared activity are known to obey a {\it
dynamical neutrality} condition. This condition enforces their sum to vanish
(over the whole finite system, or over the crystal cell, respectively). We
extend this sum rule to the non trivial case of the surface of a semiinfinite
solid and show that, in the case of a polar surface of an insulator, the
surface ions {\it cannot} have the same dynamical charges as in the bulk. The
sum rule is demonstrated through calculations for the Si-terminated SiC(001)
surface. \end{abstract}

\pacs{{\sf Preprint SISSA/14/97/CM/SC}}
]
\narrowtext

The basic quantity addressed in this work is the dynamical charge of a
given ion $s$ in different environments: within a molecule, in the bulk of a
crystalline solid, and at a solid surface. This charge is a cartesian tensor,
${\sf Z}^*_s$, which has the point symmetry of the ionic site: its components
$Z^*_{s,\alpha\beta}$ measure the dipole linearly induced (in the $\alpha$
direction) by a unit displacement of the ion $s$ (in the $\beta$
direction)~\cite{nota1}. Equivalently, ${\sf Z}^*_s$ measures the force
linearly induced on the given ion $s$ by a unit electric field (at zero
displacement): the dynamical charges govern therefore the infrared activity
of the system. The two cases of molecules~\cite{Amos87} and of bulk
solids~\cite{PCM,rap_a12} have received previous attention in the literature,
both as a matter of principle and as a subject of practical calculations,
while the case of a crystal surface has never been
considered~\cite{Ancilotto}. In a neutral molecule the sum (over all the
ions) of the dynamical charges must vanish, since a rigid translation of the
molecule as a whole induces no dipole: we will refer to this sum rule as to
dynamical neutrality. The analogue in a crystalline dielectric goes under the
name of {\it acoustic sum rule} (ASR), and is spelled out in an equally
simple manner: the dynamical charges sum to zero over the crystal cell.
However, the underlying theory is by far less trivial~\cite{PCM}.  We show
here that a crystalline surface must be dynamically neutral in order to
ensure that a rigid translation of the semiinfinite solid as a whole does not
affect the work function. Explicit formulation of such neutrality requires in
general the regularization of a nonconvergent sum. The result is a constraint
for the dynamical charges of the surface ions: in the particular case of a
polar surface---such as (001) in the zincblende structure---the novel sum
rule {\it forbids} ions of a given chemical species to have the same
dynamical charge at the surface and in the bulk.

The dynamical charges---both in molecules and in solids---have in general
nothing to do with the nominal static charge of the given ion. The difference
between dynamical and static charges is particularly dramatic when the
material has a mixed ionic-covalent character~\cite{rapix}: essentially, the
dynamical charge measures the current flowing along the bonds when the
bond-lengths are varied. In the molecular case the charge perturbation
induced by an ionic displacement is localized, and the definition of ${\sf
Z}^*_s$ is straightforward and unique~\cite{Amos87}. In a crystalline solid
one refers instead to the rigid displacement of a whole sublattice: since the
induced charge is a periodic function, one has to carefully specify the
boundary condition assumed in the solution of Poisson equation. Usually one
defines two kinds of tensors: the transverse (or Born) charge ${\sf
Z}^{*({\rm T})}_s$, and the longitudinal (or Callen) one ${\sf Z}^{*({\rm
L})}_s$. The former is the most fundamental when dealing with bulk
properties~\cite{PCM}: it is defined via the macroscopic polarization $\Delta
{\bf P}$ linearly induced in the solid by a rigid displacement of the $s$
sublattice by an amount ${\bf u}_s$, {\it while the field is kept vanishing}:
\begin{equation} \Delta P_\alpha = \frac{1}{\Omega} \sum_\beta Z^{*({\rm
T})}_{s,\alpha\beta} u_{s,\beta}, \end{equation} where $\Omega$ is the cell
volume. The longitudinal charge ${\sf Z}^{*({\rm L})}_s$ is analogously
defined, but the sublattice displacement is performed in a depolarizing field
$\Delta {\bf E} = - 4 \pi \Delta {\bf P}$.  The relationship between the two
is ${\sf Z}^{*({\rm T})}_s = \varepsilon_\infty {\sf Z}^{*({\rm L})}_s$,
where $\varepsilon_\infty$ is the electronic dielectric constant (we assume
it isotropic for the sake of simplicity). The sum rule is usually stated for
the transverse charges~\cite{PCM}, but equivalently holds for the
longitudinal charges as well. We wish to deal with molecules, surfaces, and
bulk solids all on the same footing: then the {\it longitudinal} charge is
the quantity of choice in the crystalline case.  Suppose in fact we displace
by an amount ${\bf u}_s$ only a single ion of species $s$ in an
infinite---and otherwise unperturbed---crystal: the induced charge is
localized~\cite{nota2}, and it is easy to prove that its dipole {\bf d} is
(to linear order) \begin{equation} d_\alpha = \sum_\beta Z^{*({\rm
L})}_{s,\alpha\beta} u_{s,\beta}. \label{longitudinal} \end{equation} This
can be regarded as an equivalent definition of the longitudinal charge. An
identical definition as Eq.~(\ref{longitudinal}) holds for the (unique)
dynamical charge in the molecular case. We therefore drop the superscript (L)
in the following, and we assume Eq.~(\ref{longitudinal}) as a uniform
definition of the dynamical charge for a molecule, a bulk solid, and a
surface: whenever a bulk charge is concerned, we implicitly refer to the
longitudinal one.

We are now ready to discuss the surface problem. Suppose we have a
semiinfinite crystalline insulator, where we distinguish a surface region and
a bulk region. We assume that the macroscopic field vanishes both in the bulk
of the solid, and outside in the vacuum region: this amounts to require that
no static charge is present at the surface. In these hypotheses, the average
of the electrostatic potential in the bulk of the solid with respect to the
vacuum level is a well defined quantity, which in fact determines the work
function of the given surface. We are going to impose the physical
requirement that the work function is not affected by a rigid translation of
the semiinfinite solid as a whole: a necessary condition for this to occur is
our novel sum rule for the dynamical charges.  

Suppose the surface is normal to the $z$ axis: the system has then a
two-dimensional periodicity normally to $z$: we indicate with $A$ the area of
the unit cell.  We consider a rigid displacement of an ionic layer, {\it
i.e.} we displace by an amount ${\bf u}_s$ a given ion and all its
translationally equivalent ones: this rigid displacement induces a dipole per
unit area, hence a potential lineup across the layer whose value is given by:
\begin{equation} \Delta \phi = \frac{4 \pi}{A}  \sum_\beta Z^{*}_{s,3\beta}
u_{s,\beta}.  \label{lineup} \end{equation} A rigid translation of the
semiinfinite crystal by an amount {\bf u} induces a total lineup which is, by
linearity, the sum over $s$ of the expressions in Eq.~(\ref{lineup}). As
anticipated above, we explicitly require this lineup to vanish, hence a 
na\"\i f expression for the constraint appears to be: \begin{equation} \sum_s
Z^{*}_{s,3\beta} = 0 \;\; \mbox{(any $\beta$)}. \label{srule} \end{equation}
This formal expression for the dynamical charge neutrality of the surface
cannot be used as such, given that the infinite sum in general does not
converge: in the bulk region it oscillates periodically, owing to bulk
dynamical-charge neutrality.

Therefore the problem is to regularize the indeterminate sum in
Eq.~(\ref{srule}) by using the appropriate physical criterion. The universal
panacea for this class of problems at large is the macroscopic average
introduced in Ref.~\cite{macro} and widely used by several authors: by
construction, the macroscopic average yields the correct
electrostatic-potential average in the bulk region.  Application to the
present case is straightforward.  One first maps the problem into a simple
electrostatic one by assigning a point charge of magnitude $Z^{*}_{s,3\beta}$
to each ion.  Secondly, one evaluates the planar average of this charge
distribution, which takes the general form: \begin{equation}
\overline{\rho}(z) = \frac{1}{A} \sum_s Z^{*}_{s,3\beta} \, \delta(z-z_s) ,
\end{equation} where $z_s$ are the positions of the ionic planes along the
$z$ axis. Finally, one filters $\overline{\rho}(z)$ through the convolution:
\begin{equation} \overline{\overline{\rho}}(z) = \frac{1}{b}
\int_{z-b/2}^{z+b/2} dz' \, \overline{\rho}(z'), \label{macro} \end{equation}
where $b$ is the one-dimensional periodicity of the bulk region. The result
is a piecewise constant function, vanishing both in the vacuum and in the
bulk, and whose integral in the surface region is trivial. We state the sum
rule by requiring this integral to vanish. It is important to realize
that---at variance with the  na\"\i f expression of Eq.~(\ref{srule})---the
surface sum rule involves {\it both} the values of the dynamical charges {\it
and} the coordinates of the planes $z_s$.  

The nontrivial content of the sum rule will be made clear with two examples.
We consider simple polar surfaces of a cubic binary crystal, whose bulk
dynamical charges have the form $Z^{*({\rm bulk})}_{s,\alpha\beta} = (-1)^s
|Z^*| \, \delta_{\alpha\beta}$. The $\beta \neq 3$ components of the sum rule
expressed by Eq.~(\ref{srule}) are correct, because of periodicity, for the
same reason why this equation is correct in the bulk.  We focus then on the
$\beta = 3$ component, and we further limit ourselves to the cases where the
ionic planes parallel to $z$ contain either cations only, or anions only,
alternately. This is the case for the (001) and (111) surfaces in the
zincblende structure, which we are going to illustrate separately, assuming
an ideal (truncated-bulk) geometry {\it i.e.} neglecting surface relaxation
and recostruction.

We start with the (001) surface, where $b$ is one half of the cubic lattice
constant and the ionic planes are equally spaced by an amount $b/2$.  We
assume that the crystal lies in the positive-$z$ half-space, so that the
coordinate of the $s$-th plane can be assumed to be $(s-1)b/2$.  The
piecewise constant function $\overline{\overline{\rho}}(z)$ vanishes from
$-\infty$ to $-b/2$; it assumes the value $Z^{*}_{1,33}$ (apart for a
constant factor) up to $z=0$; then $\overline{\overline{\rho}}(z) =
Z^{*}_{1,33} + Z^{*}_{2,33}$ up to $b/2$, and in general $Z^{*}_{s,33} +
Z^{*}_{s+1,33}$ in the following intervals, until it vanishes again in the
bulk region, say for $s > N$. Since the the intervals are of equal length,
the sum rule for the integral of $\overline{\overline{\rho}}(z)$ requires:
\begin{equation} 2 \sum_{s=1}^N Z^{*}_{s,33} +  Z^{*}_{N+1,33} = 0. 
\label{srule3} \end{equation} We arrive thus at an outstanding finding: the
dynamical charges of the ions in the surface region sum up to {\it one half}
of the bulk dynamical charge (with the appropriate sign). This is precisely
the relationship that one would obtain by replacing the dynamical charges
with static point-like charges and requiring the surface to be neutral.  

Switching now to the (111) surface, the linear period $b$ is $\sqrt{3}$ times
the lattice constant, and the ionic planes are no longer equally spaced: the
sum of the dynamical charges in the surface region is then a certain fraction
($\pm 1/4$ or $\pm 3/4$) of the bulk dynamical charge. For any polar surface
we have therefore a strong constraint, which {\it e.g.} forbids the surface
ions to have the same dynamical charges as in the bulk.

We illustrate our main finding using first-principles calculations of the
surface dynamical charges for a paradigmatic test case: to keep matters
simple we deal with an insulating surface. Our choice is the (001)
Si-terminated surface of the zincblende semiconductor SiC, which has been the
subject of recent theoretical work~\cite{Pollmann,Catellani}. The actual
structure is 2$\times$1 reconstructed, but even the ideal (truncated bulk)
one is insulating, and fits well our purpose of dealing with a test case as
simple as possible: the sum rule for this case takes precisely the form of
Eq.~(\ref{srule3}). All calculations are performed using density-functional
theory in the local-density approximation~\cite{DFT}. Most technical
ingredients are pretty standard: plane-wave basis sets and norm-conserving
pseudopotentials~\cite{pseudi}, Ceperley-Alder exchange-correlation
potential~\cite{XC}, and special-point Brillouin-zone sampling. We use plane
waves up to a kinetic energy cutoff of 18 Ry.  This basis set is too small if
one aims at a precise prediction of the physical properties of any
carbon-based materials, but it is enough to demonstrate the points of
principle addressed here, because the ASR in the bulk is well satisfied. We
use a set of 28 irreducible special points in the bulk calculations, and a
consistent set in the supercell ones. This large number of special points is
needed to satisfy the ASR~\cite{BGT}.  

Our calculations provide a theoretical lattice constant of 8.24 a.u. (Expt. 
8.25). Density-functional perturbation theory (DFPT)~\cite{BGT}, implemented
as in Ref.~\cite{Giannozzi91}, provides for the bulk solid $Z^{*({\rm
T})}_{\rm Si} = +2.58$, $Z^{*({\rm T})}_{\rm C} = -2.57$, and
$\varepsilon_\infty = 7.40$. The experimental estimate~\cite{Cardona} is $|
Z^{*({\rm T})}| = 2.7$. Notice that the algorithm does not enforce
neutrality: the error monitors the numerical accuracy of the calculation. The
longitudinal dynamical charge of our bulk material is therefore $|Z^*| =
0.34$.

\begin{figure}  \centerline{\psfig{file=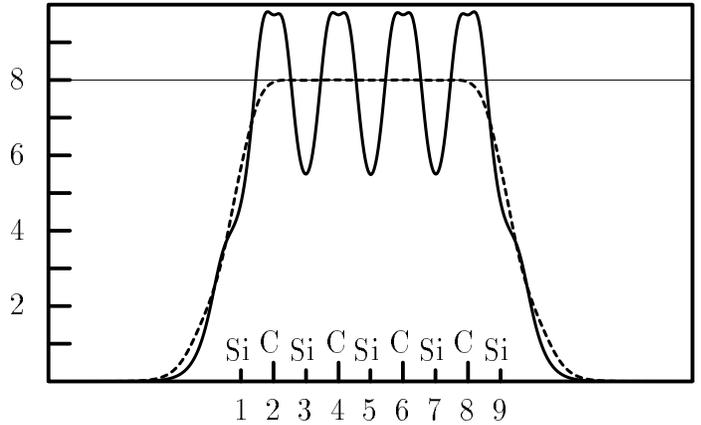,width=8cm}}
\caption[1]{ Self-consistent valence electron density in the 9-atom
computational supercell, in units of electrons per bulk cell. Vertical bars
on the abscissae indicate the positions $z_s$ of the ionic planes. The solid
line shows the average over planes parallel to the surface, as a function of
the normal coordinate $z$. The dashed line shows the same function after the
macroscopic-average filtering.}
\label{fig1} \end{figure}

The surface calculations have been performed in a supercell geometry, where
the SiC slab has a double Si termination. We plot in Fig.~\ref{fig1} the
pseudopotential valence-electron density for the 9-atom supercell; we also
plot the macroscopic average, Eq.~(\ref{macro}), of the same density. 
Fig.~\ref{fig1} clearly shows that the two surfaces are well separated, and
that the central region of the slab is bulklike, with eigth electrons per
cell in average. We have also accurately checked that each of the two
Si-terminated surfaces is statically neutral.

Starting from the reference equilibrium ground state of Fig.~\ref{fig1}, we
calculate the dynamical charges of all the ions in the supercell using again
DFPT for our composite structure of periodically repeated slabs. Obviously,
the calculated macroscopic dielectric tensor is a rather artificial quantity,
only indirectly related to genuine material properties; equally artificial
are the transverse charge tensors. As already stressed, the longitudinal
dynamical charges are the relevant physical quantities in this problem: they
are anisotropic tensors in the surface region, and they must converge to
their (isotropic) value $\pm 0.34 \, \delta_{\alpha\beta}$ in the bulk
region. The convergence proves to be rather slow: insofar as the dynamical
charges are concerned, the surface region is much larger than the equilibrium
charge density of Fig.~\ref{fig1} would suggest. Despite this fact, our slab
is large enough to recover the bulk $Z^*_{33}$ value for the center Si ion
(0.33 vs.  0.34 from the bulk calculation).  The calculated relevant
effective charges are: $Z^*_{1,33}(Si) = +0.13$, $Z^*_{2,33}(C) = -0.27$,
$Z^*_{3,33}(Si) = +0.29$, $Z^*_{4,33}(C) = -0.31$, $Z^*_{5,33}(Si) = +0.33$. 
The robustness of these figures has been then checked in two different ways. 
Firstly we have performed similar calculations on a fully relaxed (though
unreconstructed) structure: the outermost SiC bond length increases by 4\%. 
The picture is unchanged, but the bulklike limit is recovered faster, with
$Z^*_{33} = 0.34$ for the center Si ion.  Secondly the adequacy of our 9-atom
supercell has been checked against a few test calculations performed with a
slab of 13 atoms. Using the data reported above, the sum of the dynamical
charges $Z^*_{s,33}$ over five layers in the surface region, up to the
central Si, is 0.17, thus demonstrating our main finding of
Eq.~(\ref{srule3}).

A moment of reflection shows that the surface sum rule looks like equivalent
to the previously known (molecular) sum rule for the finite slab. However,
this is a very special case, where two equivalent polar surfaces are possible.
In a more general case---such as SiC(111)---terminating the bulk with two
equivalent surfaces is impossible. We have then a different sum rule for each
surface of the slab: only the sum of the two can be related to a previously
known sum rule.

In this work we have discussed the concept of ionic dynamical charge at the
surface of a crystalline dielectric. We have shown that the longitudinal
charge is the most fundamental quantity, at variance with bulk problems where
the transverse one plays the major role. A neutral surface must also be
dynamically neutral, and this imposes a nontrivial constraint for the
dynamical charges near a polar surface. We have performed a case-study
calculation for the Si-terminated SiC(001) surface. The results, besides
demonstrating the novel sum rule, also show (in this particular case at
least) that the dynamical charges converge to their bulk value more slowly
than the equilibrium electronic density would suggest.

We are grateful to S. de Gironcoli and J. Pollmann for invaluable discussions
and suggestions. Work partly supported by ONR grant N00014-96-1-0689.


\begin{references}

\bibitem{nota1} Atomic units of charge are used throughout.

\bibitem{Amos87} R.D. Amos, in: {\it Ab-Initio Methods in Quantum Chemistry -
I}, edited by K.P. Lawley (Wiley, New York, 1987), p. 99.

\bibitem{PCM} R. Pick, M.H. Cohen, and R.M. Martin, Phys. Rev. B {\bf 1}, 910
(1970).

\bibitem{rap_a12} R. Resta, Rev. Mod. Phys. {\bf 66}, 899 (1994).

\bibitem{Ancilotto}  We are aware of only one calculation, with no discussion
of the general problem:  F. Ancilotto et al., Phys. Rev. B {\bf 43}, 8930
(1991).

\bibitem{rapix} M. Posternak, R. Resta, and A. Baldereschi, Phys. Rev. B {\bf
50}, 8911 (1994); R. Resta, S. Massidda,  M. Posternak, and A. Baldereschi,
Mat. Res. Soc. Symp.  Proc. {\bf 409}, 9 (1996).

\bibitem{nota2} The induced charge is only weakly localized ({\it e.g.} not
exponentially). At large distances, it goes to zero as r$^{-3}$ times a
function which oscillates around zero with the lattice periodicity. This is a
typical ``local-field effect'' in the microscopic dielectric response of a
lattice-periodical medium.

\bibitem{macro} A. Baldereschi, S. Baroni, and R. Resta, Phys. Rev. Lett. 
{\bf 61}, 734 (1988). A thorough account is in: S. Baroni, R. Resta, A. 
Baldereschi, and M. Peressi, in: {\it Spectroscopy of semiconductor
microstructures}, edited by G. Fasol, A. Fasolino and P. Lugli, NATO ASI
Series B, vol 206 (Plenum Publishing, New York, 1989), p 251.

\bibitem{Pollmann} M. Sabisch, P. Kr\"uger, A. Mazur, M. Rohling, and J.
Pollmann, Phys. Rev. B {53}, 13121 (1996).

\bibitem{Catellani} A. Catellani, G. Galli, and F. Gygi,  Phys. Rev. Lett.
{\bf 77}, 5090 (1996).

\bibitem{DFT} {\it Theory of the Inhomogeneous Electron Gas}, edited by S.
Lundqvist and N.H. March (Plenum, New York, 1983).

\bibitem{pseudi} For C we use the same pseudopotential   
as in P. Pavone, K. Karch, O. Sch\"utt, W. Windl, D. Strauch,
P. Giannozzi, and S. Baroni, Phys. Rev. B {\bf 48}, 3156 (1993);
for Si we use the one from X. Gonze, P. K\"ackell, and M.
Scheffler, Phys. Rev. B {\bf 41}, 12264 (1990).

\bibitem{XC} D.M. Ceperley and B.J. Alder, Phys. Rev. Lett. {\bf 45}, 566
(1980); J. Perdew and A. Zunger, Phys. Rev. B {\bf 23}, 5048 (1981).

\bibitem{Cardona} D. Olego, M. Cardona, and P. Vogl, Phys. Rev. B {\bf 25},
3878 (1982). 

\bibitem{BGT} S. Baroni, P. Giannozzi, and A. Testa, Phys. Rev. Lett.
{\bf 58}, 1861 (1987).

\bibitem{Giannozzi91} P. Giannozzi, S. de Gironcoli, P. Pavone, and S. Baroni,
Phys. Rev. B {\bf 43}, 7231 (1991).

\end{references}
\end{document}